\documentclass[epsf]{mn2e}
\def\mmm{(m-M)$_0$}
\def\ebv{E(B$-$V)~}

\def\gsim{\;\lower.6ex\hbox{$\sim$}\kern-7.75pt\raise.65ex\hbox{$>$}\;}
\def\lsim{\;\lower.6ex\hbox{$\sim$}\kern-7.75pt\raise.65ex\hbox{$<$}\;}

\title[Berkeley 29]{Berkeley 29, the most distant old open cluster}

\author[Tosi et al.]{M. Tosi$^1$, L. Di Fabrizio$^2$,  A. Bragaglia$^1$, 
        P.A. Carusillo$^3$, G. Marconi$^{4,5}$  \\  
		\\
 $^1$ INAF-Osservatorio Astronomico di Bologna, Via Ranzani 1, I-40127 Bologna,
      Italy, 
      e-mail monica.tosi, angela.bragaglia @bo.astro.it \\
 $^2$ INAF-Telescopio Nazionale Galileo, 38700 Santa Cruz de La Palma, Spain,
      e-mail difabrizio@tng.iac.es  \\
 $^3$ Dipartimento Astronomia, Universit\`a di Bologna,  
      Via Ranzani 1, I-40127 Bologna, Italy \\
 $^4$ INAF-Osservatorio Astronomico di Roma, Via dell'Osservatorio 5, I-00040 
      Monte Porzio, Italy\\
 $^5$ European Southern Observatory, 
      Alonso de Cordova 3107, Vitacura, Santiago, Chile, gmarconi@eso.org
       }
\date{}

\begin{document}
\maketitle

\begin{abstract}
We present CCD BVI photometry of the old open cluster Berkeley 29, located in 
the anticentre direction. 
Using the synthetic Colour - Magnitude Diagrams technique we estimate at the
same time its age, reddening, distance, and approximate metallicity using three
types of stellar evolutionary tracks. 
The best solutions give: age=3.4 or 3.7 Gyr,
(m-M)$_0$ = 15.6 or 15.8 with  E(B--V)  = 0.13 or 0.10, and metallicity lower
than solar (Z=0.006 or 0.004), depending on the adopted stellar models. 
Using these derived values, Be 29 turns out to be the most distant open
cluster known, with Galactocentric distance R$_{\rm GC}$ = 21.4 to 22.6 kpc.

Hence, Be 29 qualitatively follows both the age--metallicity relation and the
metal abundance gradient typical of Galactic disc objects. The cluster position and 
radial velocity, however, appear to link Be 29 to the family
of the Canis Major debris. 

\end{abstract}


\begin{figure*}
\vspace{13cm}
\includegraphics{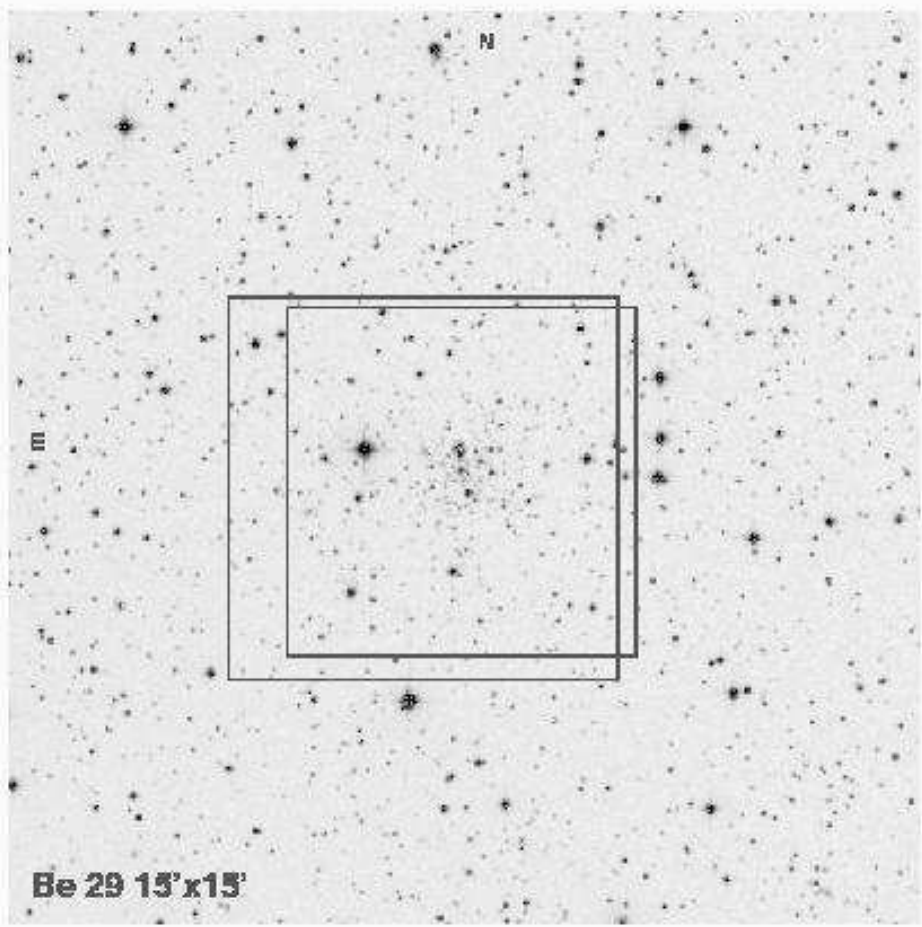}
\caption{Position of our fields: the slightly larger one is for the Danish
observations, the other for the SuSI2 ones (notice that in this case
there is a small vertical gap between the 2 CCDs); the map is 15 x 15 arcmin, 
and is oriented with North up and East left.} 
\label{fig-map}
\end{figure*}

\begin{figure*}
\vspace{13cm}
\includegraphics{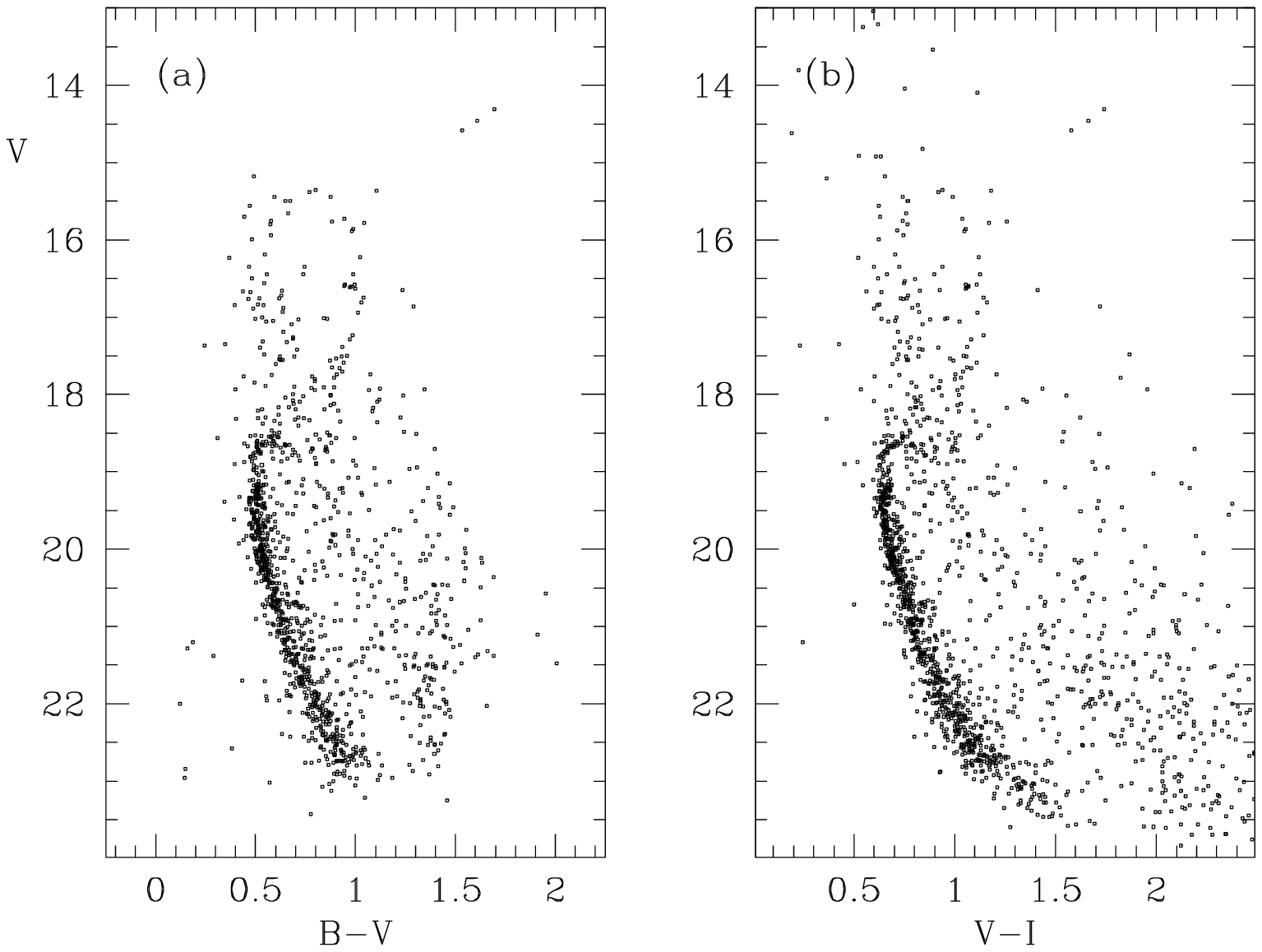}
\caption{CMDs of the Be 29 field based on our photometry, combining the Danish
and SuSI2 exposures. Note the well populated main sequence of the cluster, easy
to distinguish from the field stars one, which is redder and with a much
brighter TO (around V ~$\simeq$ 15.5). The RGB and red clump regions are also
clearly visible, but more difficult to disentangle from the field component.} 
\label{fig-cmd}
\end{figure*}

\section{Introduction}

As amply acknowledged (e.g. Friel 1995), open clusters (OC's) are among the
best objects to describe the Galactic disc properties. In particular,
old OC's may be used to trace the disc formation and evolution,
and the history of its chemical enrichment. There is a metallicity
gradient in the Galactic disc, with the inner regions being more metal
rich than the outer ones, as established on the basis of different indicators, 
from HII regions (e.g. Shaver et al. 1983) and young B stars (e.g. Smartt 
\& Rollerston 1997), to planetary nebulae (e.g. Pasquali \& Perinotto 1993).
Open clusters have been suggested to show a metallicity gradient as well
(e.g. Janes 1979, Panagia \& Tosi 1981, Friel  1995, Carraro, Ng \& Portinari 
1998, Friel et al. 2002), although  Twarog et al. (1997) argue instead that
what is observed is a step distribution in OC metallicity. 

While there is quite
a wealth of OC's with known metallicity in the Sun vicinity, only few objects
farther away have been measured: for instance, in Friel's (1995) compilation
there are only two OC's with Galactocentric distance larger
than 15 kpc (Berkeley 20 and Berkeley 29, this last being the farthest known
OC), and another one is present in Twarog et al. (1997: Tombaugh 2). 
Finding and studying clusters at large distances from the Sun, both towards the
Galactic centre and the anticentre is essential to define how the disc
properties vary radially. Recently Frinchaboy \& Phelps (2002) suggested  that
the most distant old open cluster is  Saurer A (with Galactocentric distance of
more than 19 kpc);  Carraro \& Baume (2003) confirmed their results, finding an
age of about 5 Gyr, Z $\sim$ 0.008, and distance from the Sun of 13.2 kpc,
implying a Galactocentric distance of about 21 kpc. Even more recently, all
these old and distant OC's, and in particular Saurer A and Be 29, have been
proposed to be originated in/by the Canis Major dwarf spheroidal galaxy, the
satellite which is merging with the Milky Way near the Galactic plane (Martin
et al. 2004, Bellazzini et al. 2004, Frinchaboy et al. 2004).

This paper is part of a long term project dedicated to the study of (mostly
old) OC's both with precision photometry (e.g. Bragaglia \& Tosi 2003,
Andreuzzi et al. 2004, Kalirai \& Tosi 2004, and references therein)
and with high resolution spectroscopy
(Bragaglia et al. 2001, Carretta et al. 2004).
The old open cluster Berkeley 29 (C0650+169, OCL486) has $\alpha_{2000} =
06^h53^m04^s$, $\delta_{2000} = +16^o 55 \arcmin 39 \arcsec$, l= 197.98, b=8.03. 
We chose to observe it because it appeared to be the most distant open cluster
known, and to have old age and low metallicity (Kaluzny 1994). 
Even if Saurer A were $the$ most distant OC, 
Be 29 would retain its importance, since it is more populous; moreover, 
as we will show in this paper, the distance to Be 29 has been underestimated
in the past and we do find it to be slightly more distant than Saurer A.
For these
reasons Be 29 is of  paramount importance in defining the disc abundance
gradient (if its origin is completely Galactic) or the properties and
formation history of
accreted clusters (if it originates from the merged satellite). 

Perhaps due to its faintness, Be 29 has not been the subject of extensive
works in the past: the first and only calibrated colour-magnitude diagram
published to date is the one by Kaluzny (1994, hereafter K94).
K94 observed in the B, V, and I bands a small field (3 arcmin $\times$ 
8 arcmin) with the 2.1m KPNO telescope, and presented deep and well defined 
CMDs. From the projected density distribution of stars he derived a
radius of about 1.5 arcmin ~on the sky. 
He estimated the following properties: 
age $\simeq$ 4 Gyr, distance from the Sun d = 10.5 kpc, Galactocentric
distance R$_{\rm GC}$ = 19 kpc, and low metallicity [Fe/H] $\lsim$ --1.
To derive these figures he assumed \ebv = 0.21 from Burstein \& Heiles (1982).
Note however that the new maps of interstellar absorption by Schlegel,
Finkbeiner \& Davis (1998), appropriate for b $>$ 5, give the much lower value
\ebv = 0.093, which would somewhat alter those figures (towards higher
metallicity and larger distance from the Sun).

Phelps, Montgomery \& Janes (1994, hereafter PJM94) observed the cluster,
but published only non calibrated CMDs (instrumental B, V, I magnitudes).
They derive an age of 2.1 Gyr, based on the difference in magnitude between
the red clump and the main sequence turn-off, and do not give any value
for reddening and metallicity. In a companion paper (Janes \& Phelps 1994),
a distance of 8.6 kpc from the Sun is assigned to Be 29, but based on the
assumption of a "mean" diameter of 5 pc for the old cluster population, hence
not very significant.

Noriega-Mendoza \& Ruelas-Mayorga (1997) applied to Be 29 a
technique for the simultaneous determination of metallicity and reddening,
similar to the one developed by Sarajedini (1994) for globular clusters,
useful in stellar systems with well defined giant branch and helium burning
clump. At variance with the suggestion by K94, they derived [Fe/H]=$-0.30$, and
\ebv=0.01; their estimated uncertainties on these values are $\pm$0.04 in
[Fe/H] and $\pm$0.02 in \ebv. These results are at odds with what has
been obtained by all other studies.

Recently, we  have obtained intermediate resolution spectra of 20 stars in the
direction of the cluster, and have been able to confirm membership for 12
objects in the crucial evolutionary phases of the red giant branch (RGB) and 
red clump
(Bragaglia, Held \& Tosi 2004). This information will be used in the present
analysis. Furthermore, comparison of the confirmed members to other well
studied clusters has permitted to derive indication for reddening, metallicity
(and distance) much more consistent with K94 than with Noriega-Mendoza \&
Ruelas-Mayorga (1997).

We will present our photometric data and the reduction procedure in Section 2;
the CMD, the presence of binaries and the field contamination  will be
discussed in Section 3. In Section 4 results for age, reddening and distance 
based on the synthetic CMD technique will be presented, and a summary of 
results and discussion will be given in Section 5.

\begin{figure}
\vspace{14cm}
\includegraphics{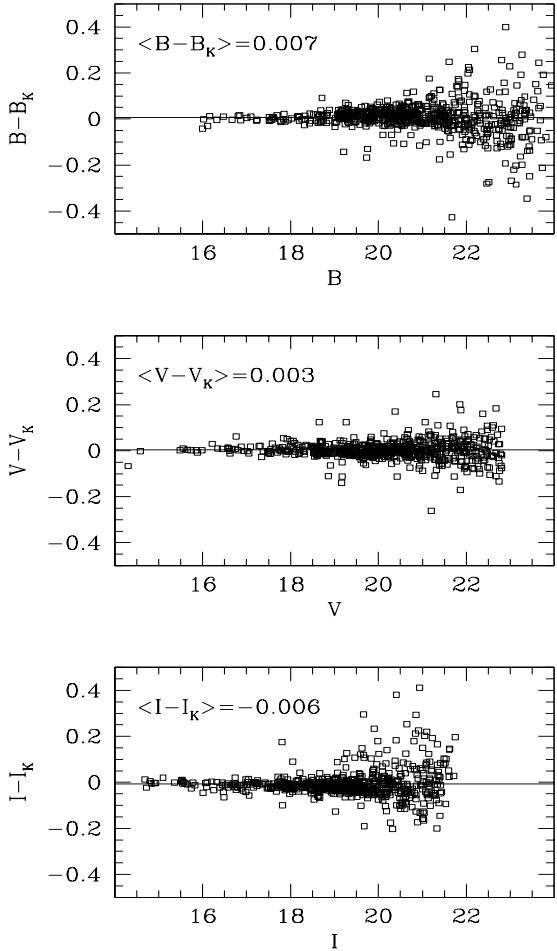}
\caption{Comparison of our B, V, I photometry with the one in Kaluzny (1994).} 
\label{fig-conf}
\end{figure}

\section{Observations and data reduction}

Be 29 was observed  on the night of 9 March 1995 with the  1.54m 
Danish telescope (La Silla, Chile) and a direct camera, mounting the CCD \#28,
a chip Tek 1024 $\times$ 1024 pixel, with scale 0.377 arcsec/pix, and field of
view of  6.4 x 6.4 arcmin$^2$. The night was part of a run
dedicated to open clusters; sky conditions were photometric, and 
seeing varied from 0.83 arcsec to 1.02 arcsec (see Table 1 for a log of
the observations). We observed a single field (see Fig. 1), 
centered on Be 29; given the
small size of the cluster, we did not obtain a separate pointing for field
stars decontamination.

To reach fainter magnitudes, these data were supplemented with new ones taken on
16-17 January 2002 with the NTT (La Silla), mounting SuSI2, the Super Seeing
Imager, which uses two chips (CCDs \#45 and 46,  EEV44-80, 2048 $\times$
4096 pixel) with a scale 0.161 arcsec/pixel and total field of view of 5.5
$\times$ 5.5 arcmin$^2$, with a small vertical gap of 8 arcsec at the
junction. In this case seeing varied from 0.66 arcsec to 1.16 arcsec.

Data reduction was done using the usual IRAF\footnote{IRAF is distributed by
the NOAO, which are operated by AURA, under contract with NSF} routines to
perform bias subtraction, flat fielding correction, and cleaning of cosmic
rays hits.  The SuSI2 I images were also corrected for fringing.

We then applied the procedure for PSF study and fitting available in
DAOPHOT--II, also in IRAF environment  (Stetson 1987, Davis 1994).
The building of the initial catalogue was slightly different for the two data
sets:  we  searched the Danish frames independently with DAOFIND and a
threshold of 3 $\sigma$  over the local sky value, while for the SuSI2 frames we
used STARFINDER (Diolaiti et al. 2000), which is very powerful in finding all 
stars, excluding at the same time false identifications (i.e. extended
objects, blends, cosmic rays, defects, etc) but is also quite slow. 
The following steps
were identical, except that the two SuSI2 chips were analyzed separately.

About 20 well isolated, bright stars were used in each frame  to define the
best analytical PSF model, which was then applied to all the
detected objects. The
resulting magnitude file was selected both in magnitude, to avoid saturated
stars, and in sharpness, a shape - defining parameter, to avoid cosmic rays and
false identifications of extended objects (this was relevant mostly for the
Danish data).

All output catalogues were aligned in coordinates, and  "forced" on a
reference frame for each filter applying a "zero point" shift to the
instrumental magnitude, using dedicated programs developed at the Bologna
Observatory by P. Montegriffo. Special care had to be taken in aligning
the SuSI2 B magnitudes, for which we had to take into account a colour term.

We computed a correction to the PSF derived magnitudes to be on the same system
as the photometric standard stars, performing aperture photometry on a few
isolated stars in the  reference images (three Danish frames, since they were
obtained in photometric conditions).  The correction to be applied to the PSF
magnitudes was found to be: --0.233 in $B$, --0.204 mag in $V$, and --0.207 in
$I$ (in the sense aperture minus PSF).
The final magnitude catalogue in each band is the result of the
(weighted) average of all measures for each stars.

The conversion from instrumental magnitudes to the Johnson-Cousins
standard system was obtained using 
the same equations already derived for this observing run in the case
of Pismis 2 (Di Fabrizio et al. 2001, to which we refer for details):

$$ B = B +0.187 \cdot (b-v) -7.326 ~~~~(r.m.s.=0.014) $$
$$ V = V +0.037 \cdot (b-v) -6.693 ~~~~(r.m.s.=0.012) $$
$$ V = V +0.032 \cdot (v-i) -6.646 ~~~~(r.m.s.=0.014) $$
$$ I = I -0.018 \cdot (v-i) -7.528 ~~~~~(r.m.s.=0.016) $$

where $b,v,i$ are instrumental magnitudes, while $B,V,I$ are the
corresponding Johnson-Cousins magnitudes. 
We calibrated the $B$ magnitudes using the relation involving
$(b-v)$,  and the $I$ and $V$ magnitudes with the ones involving $(v-i)$
(except for the stars missing $i$, for which we used the
$(b-v)$ colour to  calibrate $V$).

We tested the completeness of our luminosity function in the $B, V$ and $I$ 
band on the deepest images, i.e., the SuSI2 ones, adding artificial stars to the
frames  and 
repeating the procedure of extraction of objects
and PSF fitting used for the original frame. 
Stars were added at random positions 
and selected in magnitude according to the observed
luminosity  function, about 120 at a time, 
in order not to significatively alter the crowding conditions, and repeating 
the process until a total of about 50,000 artificial stars was reached. 
To the output catalogue of the added stars we applied the same selection 
criteria
in magnitude and sharpness as done for the science frames.
The completeness degree of our photometry at
each magnitude level was computed  as the ratio of the number of
recovered stars to the number of simulated ones  (considering as
recovered objects only those found within 0.5 pix of the given
coordinates, and with magnitudes differing from the input ones less than
$\pm$ 0.75 mag), and is given in Table~\ref{tab-compl}.

\begin{table}
\begin{center}
\caption{Log of the observations. }
\begin{tabular}{lllll}
\hline
Telescope &Date             &Exp B &Exp V &Exp I\\
          &                 &(s)   &(s)   &(s)\\
\hline
Danish    &Mar 9, 1995  & 1200        &20,60,900    &20,60,900   \\
NTT       &Jan 16, 2002 & 1800        &2$\times$900      &300,2$\times$900 \\
          &Jan 17, 2002 & 1	      &900	    &		 \\
\hline
\end{tabular}
\end{center}
\label{tab-log}
\end{table}

\begin{table}
\begin{center}
\caption{Completeness of our photometry in the three filters. }
\begin{tabular}{rrrrrr}
\hline
B &compl$_{\rm B}$ & V & compl$_{\rm V}$ & I &compl$_{\rm I}$ \\
\hline
$<$16.75 & 1.000 &$<$16.65 &1.000 &$<$15.30 & 1.000 \\
17.25	 & 1.000 &17.15 &0.968 &15.80 & 1.000 \\
17.75	 & 0.991 &17.65 &0.987 &16.30 & 1.000 \\
18.25	 & 0.994 &18.15 &0.974 &16.80 & 0.994 \\
18.75	 & 0.987 &18.65 &0.974 &17.30 & 0.995 \\
19.25	 & 0.983 &19.15 &0.971 &17.80 & 0.989 \\
19.75	 & 0.978 &19.65 &0.967 &18.30 & 0.985 \\
20.25	 & 0.980 &20.15 &0.963 &18.80 & 0.986 \\
20.75	 & 0.974 &20.65 &0.946 &19.30 & 0.975 \\
21.25	 & 0.969 &21.15 &0.938 &19.80 & 0.971 \\
21.75	 & 0.954 &21.65 &0.932 &20.30 & 0.961 \\
22.25	 & 0.949 &22.15 &0.912 &20.80 & 0.932 \\
22.75	 & 0.915 &22.65 &0.858 &21.30 & 0.859 \\
23.25	 & 0.829 &23.15 &0.753 &21.80 & 0.563 \\
23.75	 & 0.608 &23.65 &0.428 &22.30 & 0.166 \\
24.25	 & 0.271 &24.15 &0.081 &22.80 & 0.250 \\
24.75	 & 0.014 &24.65 &0.002 &23.30 & 0.000 \\
25.25	 & 0.000 &25.15 &0.000 &23.80 & 0.000 \\
\hline
\end{tabular}
\end{center}
\label{tab-compl}
\end{table}

\section{The Colour - Magnitude Diagram}

The final catalogue contains 1649 stars, of which 1144 have $B$, $V$, and $I$
magnitudes, 1182 have at least $B$ and $V$
magnitudes, and 1611 have at least $V$ and $I$ magnitudes. 
The pixel positions
of all objects were transformed to equatorial coordinates using software
written by P. Montegriffo at the Bologna Observatory, and using the Guide Star
Catalogue 2 (GSC2) as reference frame; residuals
of the transformation between the two systems (as deduced from the stars in
common) are of 0.14 arcsec in right ascension and 0.10 arcsec in 
declination.
Tables containing the photometry, the pixel and equatorial coordinates will
be available in electronic form  through the BDA\footnote{
http://obswww.unige.ch/webda/webda.html} (Base Des Amas, Mermilliod 1995).

The resulting CMDs are shown in Fig.~\ref{fig-cmd}:
the cluster sequences are very well delineated, with a main sequence (MS)
extending from $V\simeq  18.6$ and $B-V \simeq 0.45$ or $V-I \simeq 0.60$ 
at the turn-off (TO) point 
down to $V \simeq$ 23 or 23.5, for about 4 - 5 magnitudes. The subgiant
and red giant branches (SGB, RGB) are also visible, even if suffering more
from  field contamination (the Galactic disc), which appears like
a scattered main sequence, positioned mostly on the red side
of the cluster MS, and
extending to redder colours and to luminosities much brighter than the MS TO
of Be 29.
The red clump, i.e., the locus of core-He burning stars, is present at
$V \simeq 16.6$  and $B-V \simeq 0.98$, $V-I \simeq 1.1$.
Three stars stand up, at $V \simeq 14.5$, $B-V \simeq 1.6$, $V-I \simeq 1.6$:
as already noted by K94, they seem to represent the brighter part of the RGB, 
if not its tip. 

We also identified stars in common between our sample and K94, and a comparison
of the two photometries is shown in Fig. ~\ref{fig-conf}: they appear to be in
perfect agreement, with average zero point shifts of much less than 0.01 mag in
all the three filters. 

\begin{figure}
\vspace{6cm}
\includegraphics{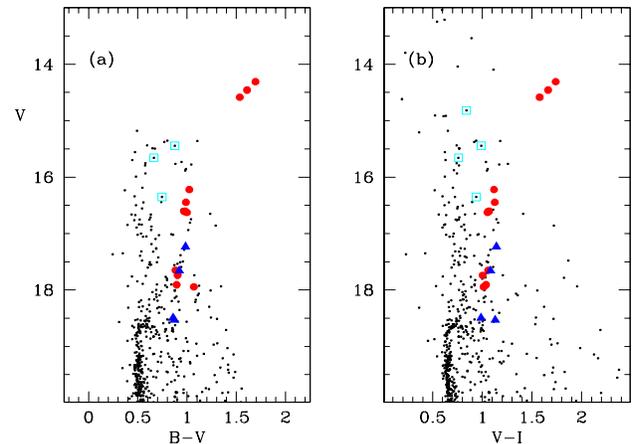}
\caption{CMDs where stars with membership information
are marked with different symbols: filled dots are members, triangles
are uncertain attributions, and open squares are field stars (taken from 
Bragaglia et al. 2004).} 
\label{fig-mem}
\end{figure}

Recently Bragaglia et al. (2004) derived radial velocities to determine
membership for a subsample of 20 stars, including the three bright RGB ones:  4
were  definitely found to be field objects, 4 have uncertain status, and 12
appear to belong to the cluster RGB and clump. These stars are indicated with
different symbols in Fig. ~\ref{fig-mem}

\begin{figure*}
\vspace{12cm}
\includegraphics{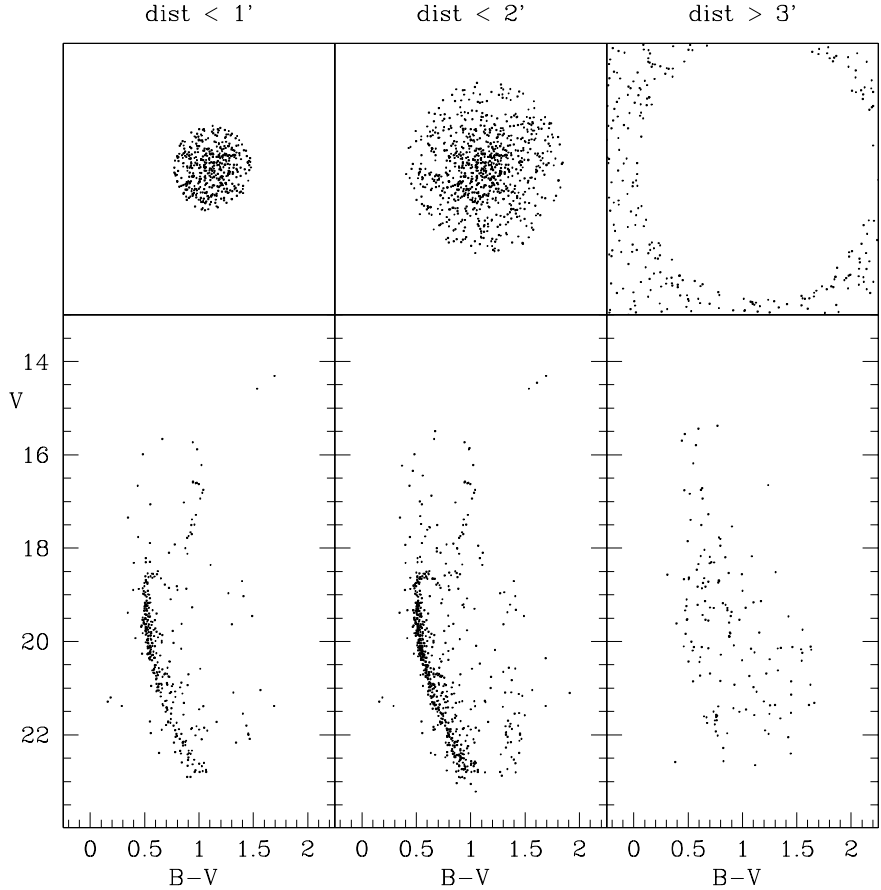}
\caption{$V, B-V$ CMDs in different zones of our field. From left to right: 
only the inner 1 arcmin
radius, the inner 2 arcmin, and the external part, the best approximation we
have of a comparison field. Be 29 appears very concentrated, even if a few
main sequence stars are present also in the farthest regions. We can easily
discriminate between the cluster and the fore/background.} 
\label{fig-rad}
\end{figure*}

This information is not sufficient to define a clean and safe cluster sample, 
so to further check on the field fore/background contamination we have plotted 
the CMDs, selecting stars near the cluster centre or far from it (see
Fig.~\ref{fig-rad}): the  cluster sequences are even easier to locate, and this
will be used in next Section, when selecting the sample of stars to simulate.

\begin{figure*}
\vspace{6cm}
\includegraphics{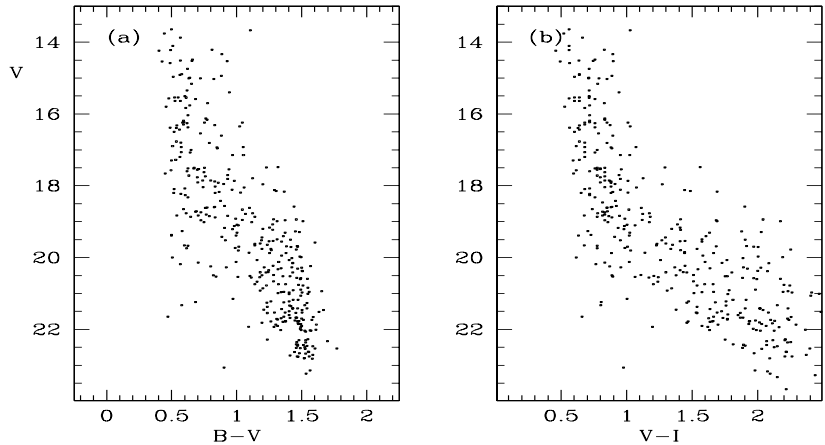}
\caption{$V, B-V$ and $V, V-I$ CMDs obtained using the Besan\c con model for
the Galaxy, useful to understand the fore/background contamination.} 
\label{fig-model}
\end{figure*}

To further test the identification of cluster and field components, we have
also used the Besan\c con models for the Galaxy stellar structure 
(Robin et al. 2003, available
at the web site {\tt www.obs-besancon.fr/www/modele/modele\_ang.html}).
Results of a query on an area equivalent to our field of
view located at l=200, b=8 are shown in Fig. ~\ref{fig-model}, where we
have applied our same incompleteness factors. It's easy to
appreciate the similarity of the model to what we have defined as field
population in our data.
In other words, we are seeing the ''normal" components of the
Milky Way, without indications of extra contribution from e.g. the
Canis Major galaxy, seen in the background of other OC's (Bellazzini et al.
2004). On the other hand, our field of view is very small (about 0.011 square
degrees) and Be 29 is not near the supposed nucleus of the disrupted
satellite (located at galactic coordinates l,b $\sim 240,-7$).

\begin{figure*}
\vspace{9cm}
\includegraphics{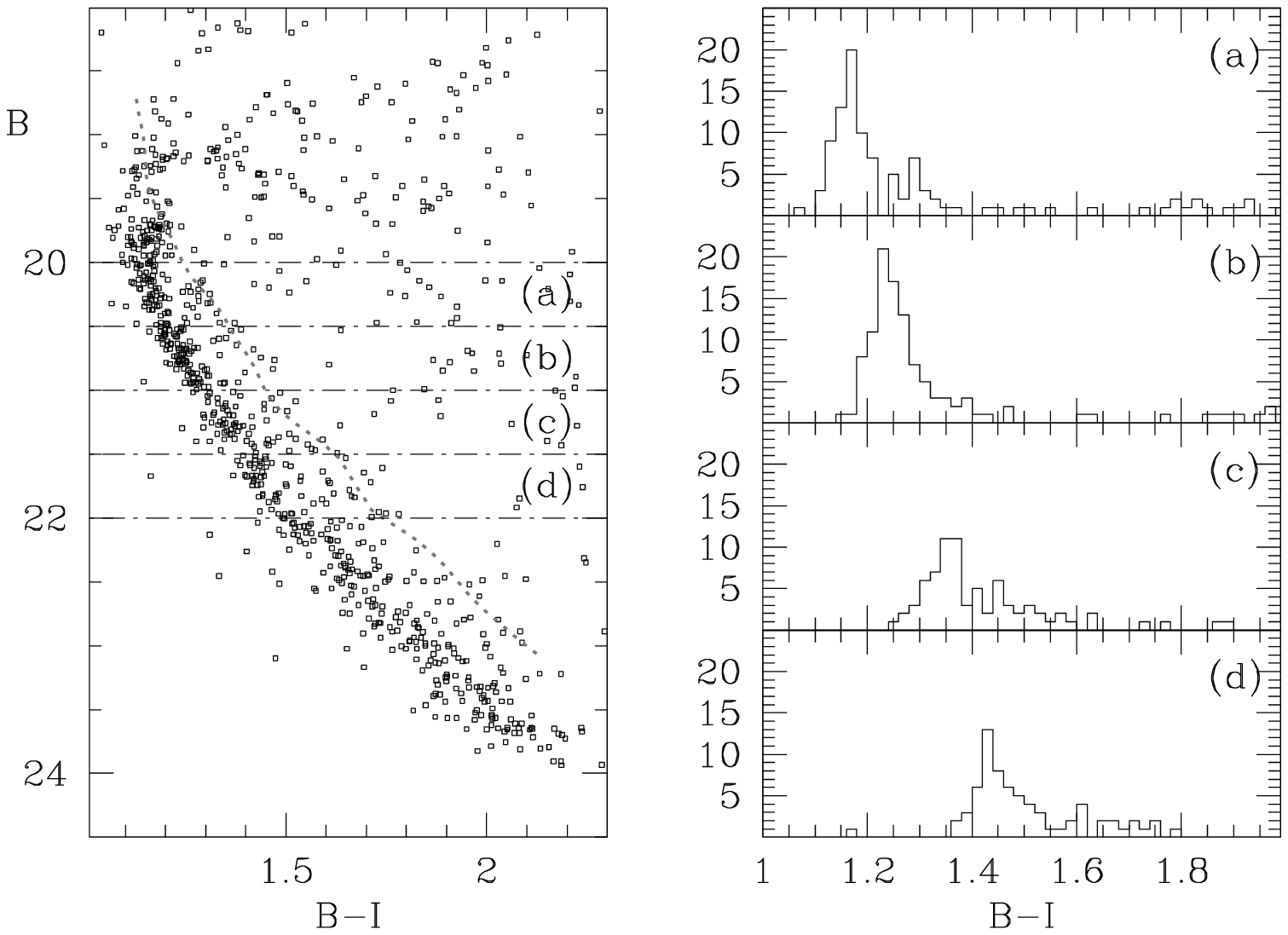}
\caption{Left panel: enlargement on the main sequence of the $B, B-I$ CMD.
The line above the MS is simply the MS ridge
line shifted by 0.75 mag, to indicate the position of the equal-mass binaries.
Right panels: histograms in colour of stars in bins of 0.5 mag from $B$=20 to 
$B$=22; the indication of a secondary peak, indicative
of binary systems, is well visible in the upper panel, while its presence is
more dubious, due to the smaller numbers, in the others.} 
\label{fig-bin}
\end{figure*}

Even if not so evident from Figs.~\ref{fig-cmd} and \ref{fig-rad}, a population
of binary systems is present in Be 29, as in all the other open clusters
examined to date. Fig.~\ref{fig-bin} shows an enlargement of the MS, plotting
the $B, B-I$ CMD, which allows for a better discrimination of the single and
binary stars main sequences. The fraction of binary systems will be
derived together with the cluster parameters in the next Section.

\begin{figure*}
\vspace{19cm}
\includegraphics{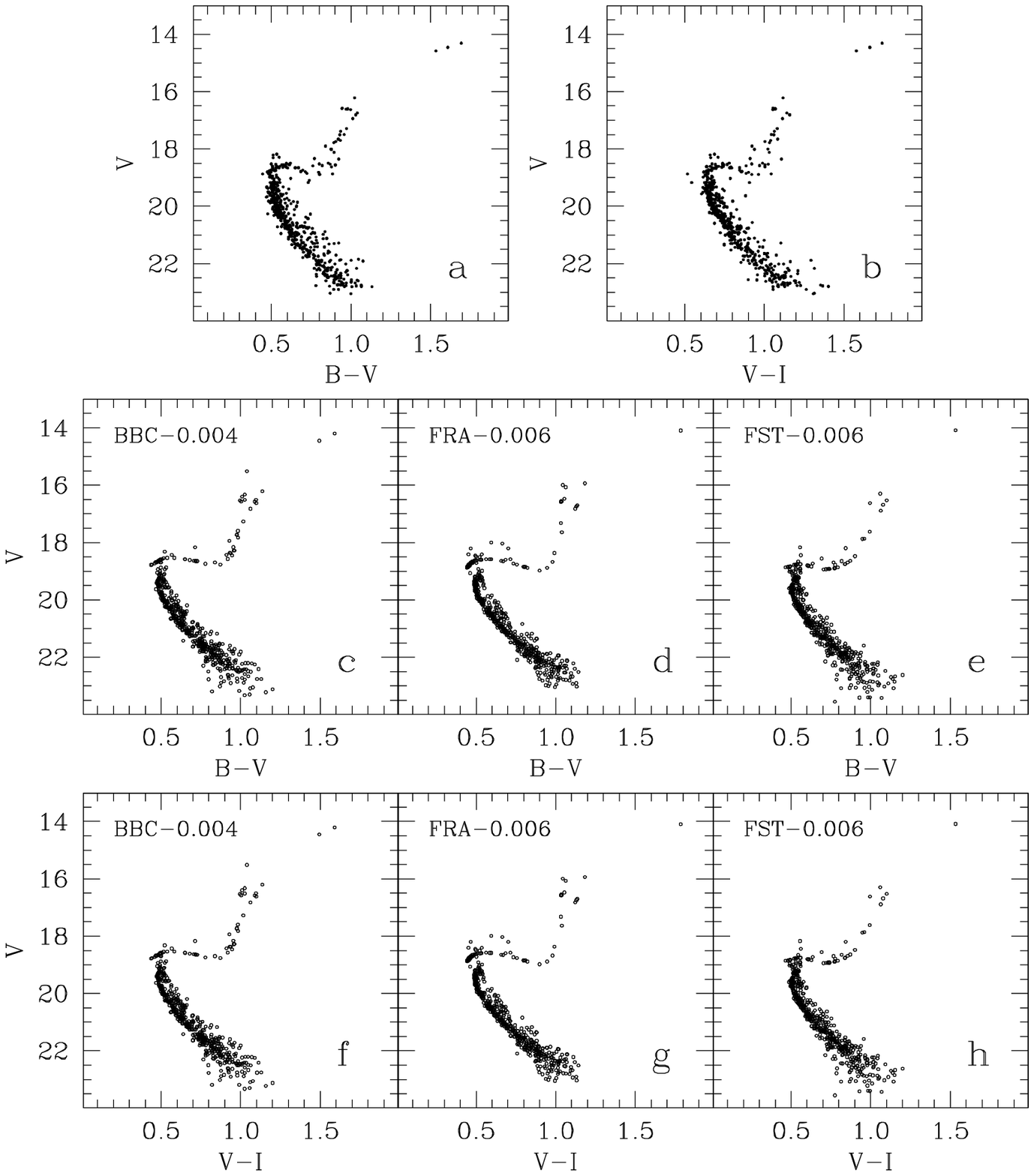}
\caption{Top panels: empirical CMDs of the region of Be 29 within 2 arcmin
from the cluster centre cleaned from likely field stars. 
Panels c through h: corresponding
synthetic CMDs in better agreement with the data for the three
different types of stellar models (labelled on the top left corner of each
box together with the model metallicity). The BBC model of panels c and f 
assumes age = 3.7 Gyr, \ebv=0.13, \mmm=15.6. The FRA model of panels d and g 
assumes age = 2.3 Gyr, \ebv=0.15, \mmm=15.8.The FST model of panels e and h 
assumes age = 3.5 Gyr, \ebv=0.10, \mmm=15.8.} 
\label{sim}
\end{figure*}

\section{Cluster parameters}

Age, distance and reddening of Be 29 have been derived with the same procedure 
applied to all the clusters of our project (see Andreuzzi et al. 2004, Kalirai
\& Tosi 2004 and references therein), namely the synthetic CMD method
originally described by Tosi et al. (1991). The best values of the parameters  
are found by selecting the cases providing synthetic CMDs with morphology, 
colours, number of stars in the various evolutionary phases and luminosity     
functions (LFs) in better agreement with the observational ones.

As usual, to test the effect of different input physics on the derived  
parameters, we have run the simulations with three different types of stellar   
evolutionary tracks, assuming different prescriptions for the treatment of 
convection and ranging from no overshooting to  high overshooting from 
convective regions. 

Be 29 is known to be rather metal poor (see the Introduction), 
but to avoid biases in the parameter determinations, we have created the 
synthetic CMDs adopting, for each type of stellar models, metallicities 
ranging from solar down to 1/20 of solar. Notice that we consider as
solar metallicity tracks those with Z=0.02, because they are the ones 
calibrated by their authors on the Sun, independently of the circumstance 
that nowadays the actual solar metallicity is supposed to be lower (see 
Asplund et al. 2004).
We recall that the position of a stellar model in the CMD depends only
on its mass, age and metallicity, but the formal effect of metallicity actually
includes that of the opacities adopted in the stellar evolution models. 
Metallicities attributed to clusters via comparison with stellar
models or isochrones therefore are always to be taken with caution. This is why
we consider only indicative the metallicities obtained with our photometric 
studies and prefer to use high resolution spectroscopy for a safe determination
of the chemical abundances.

The adopted sets of stellar tracks are listed in Table 3, where the 
corresponding references are also given, as well as the  model metallicity and
the information on their corresponding overshooting assumptions. 
The transformations from the theoretical luminosity and effective temperature 
to the Johnson-Cousins magnitudes and colours have been performed using Bessel, 
Castelli \& Pletz (1998) conversion tables and assuming E(V-I) = 1.25 \ebv
(Dean et al. 1978) for all sets of models. Hence, the different results
obtained with different stellar models must be ascribed fully to the models
themselves and not to the photometric conversions.

\begin{table}
\begin{center}
\caption{Stellar evolution models adopted for the synthetic CMDs;
the
FST models actually adopted here are an updated version of the published ones
(Ventura, private communication)}
\vspace{5mm}
\begin{tabular}{cccl}
\hline\hline
   Set  &metallicity & overshooting & Reference \\
\hline
BBC & 0.02 & yes &Bressan et al. 1993 \\
BBC & 0.008& yes &Fagotto et al. 1994 \\
BBC & 0.004& yes &Fagotto et al. 1994 \\
FRA & 0.02 & no &Dominguez et al. 1999 \\
FRA & 0.01 & no &Dominguez et al. 1999 \\
FRA & 0.006 & no &Dominguez et al. 1999 \\
FRA & 0.001 & no &Dominguez et al. 1999 \\
FST & 0.02 & $\eta$=0.02 &Ventura et al. 1998\\
FST & 0.02 & $\eta$=0.03 &Ventura et al. 1998\\
FST & 0.01 & $\eta$=0.02 &Ventura et al. 1998\\
FST & 0.01 & $\eta$=0.03 &Ventura et al. 1998\\
FST & 0.006 & $\eta$=0.02 &Ventura et al. 1998\\
FST & 0.006 & $\eta$=0.03 &Ventura et al. 1998\\
\hline
\end{tabular}
\end{center}
\label{models}
\end{table}

The synthetic stars are attributed the photometric error derived from 
the artificial stars tests performed on the actual images. They are
retained in (or excluded from) the synthetic CMD according to the photometry 
completeness  factors listed in Table \ref{tab-compl}.
All the synthetic CMDs have been computed either assuming that all the cluster
stars are single objects or that a fraction of them are members of binary 
systems with random mass ratio. We find, as in many other clusters, that a 
binary fraction around 30\% well reproduces the observed distribution along the
main sequence. All the synthetic CMDs shown in the figures assume this 
fraction of binaries. 

Membership to the cluster has been proven (Bragaglia et al. 2004) for 12 stars, 
including the three brightest red giants of the CMD. We have run the
simulations both for the CMD corresponding to the whole field covered by our
images (using the more external part of the field to estimate back/foreground
contamination) and to the inner cluster region, practically unaffected by
contamination. The results are absolutely consistent with each other and
therefore we present here only the cases corresponding to the inner cleaner
region. The examined field has a radius of 2 arcmin from the cluster centre.
After removing the stars that, on the basis of the radial velocities and of the
external field CMD, appear not to be cluster member, this field
contains 553 stars measured in all the $B$, $V$ and $I$ filters. The
corresponding CMDs are shown in  Fig.~\ref{sim} a and b. The synthetic CMDs
 have therefore  been created with this number of objects.

We find that in all cases a solar metallicity must be excluded for Be 29, since
all the synthetic CMDs with Z=0.02 are too red and we would need a non physical
negative reddening to account for the observed colours. 
Stellar tracks with metallicity
lower than solar can reproduce the observed MS colours, if an appropriate
reddening is adopted, but only those with Z = 0.004 -- 0.006 are able to 
reproduce both in $B-V$ and in $V-I$ the colours of the observed RGB. All the 
other metallicities lead to excessively red RGBs either in one colour or in 
both. Moreover, metallicities outside the indicated range sometimes lead also
to less consistent features in the CMD sequences.

More in detail, the BBC models with Z=0.004, age=3.7 Gyr, \ebv = 0.12 and
\mmm=15.60 reproduce quite well both the morphology and the number of objects 
of the CMD evolutionary sequences (MS, SGB, RGB and clump), as 
well as their colours, both in $B-V$ and in $V-I$ (see Fig.~\ref{sim} c and f).
The synthetic sequences are slightly tighter than the empirical ones, but this
may be due to residual contamination from back-foreground stars (see also the
LFs below). This aspect is found also for all the types of examined models.

The FST models with Z=0.006 and moderate overshooting reproduce equally well
the data, for age=3.4--3.5 Gyr, \ebv = 0.10 and \mmm=15.80 (see Fig.~\ref{sim} e
and h). 
They need a slightly younger age than the BBC models because they have a 
somewhat lower overshooting, and a slightly lower reddening because they are 
slightly more metal rich. The FST models with same Z=0.006 metallicity but
higher overshooting also reproduce the observed colours for an age 3.5--3.8
Gyr and the same \ebv = 0.10 and \mmm=15.80, but the morphology of the upper MS,
just below the turn-off point, is rounder than observed.

The BBC models with Z=0.008 reproduce fairly well the CMD morphology and number
counts, the best case having age=3.5 Gyr, \ebv = 0.08 and \mmm=15.75. 
However, their RGBs always turn out to be too red in $V-I$ and we then
consider this (formal) metallicity less adequate than Z=0.004.
Similarly, the FRA models with Z=0.006 reproduce rather well the cluster 
CMD and LF, but their RGB in $V-I$ are redder than observed. For these FRA
models, the best case assumes age=2.5 Gyr, \ebv = 0.15 and \mmm=15.80
(Fig.~\ref{sim} d and g). This age
is much younger than that of the BBC and FST models because 
stars of the same mass are fainter than in the overshooting case.

Also the FST tracks with Z=0.01 lead to RGBs too red in $V-I$. In addition they
have too flat subgiant branches and excessive curvature of the upper MS.
The FRA models with Z=0.001 have RGB and clump excessively red both in B--V and
in $V-I$ and a non satisfactory shape of the turn-off. The FRA models with
Z=0.01 also have excessively red RGB and clump and the tendency to
over populate them. 

\begin{figure}
\vspace{10cm}
\includegraphics{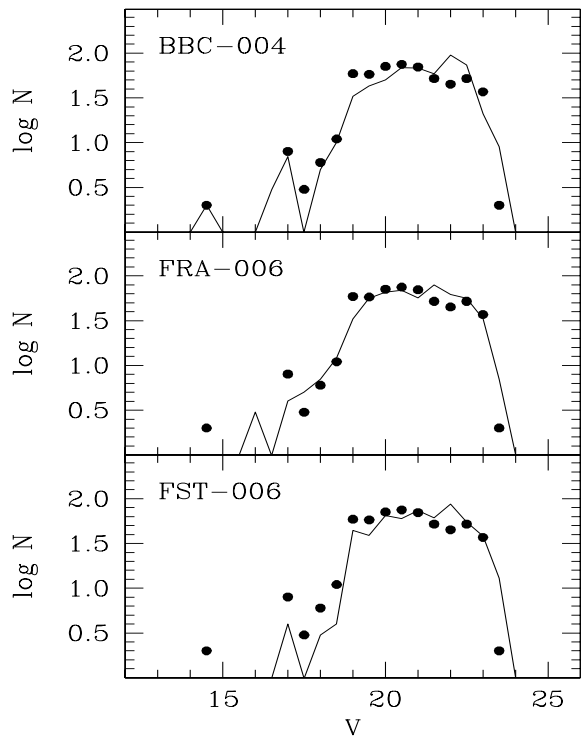}
\caption{Luminosity functions of the selected stars in the cluster region 
within 2 arcmin
from the center (dots). The lines correspond to the best synthetic models
for each type of tracks, whose CMDs are shown in Fig.\ref{sim}.  } 
\label{simlf}
\end{figure}

The LFs of the best cases for the BBC, FRA and FST models are shown in
Fig.~\ref{simlf} (lines) and compared to that of the CMD of Fig.~\ref{sim}a,
assumed to be the empirical one (dots).
In no case the fit to the data is really good: the models tend to underestimate
the number of stars brighter than $V = 20$ and to overestimate the number of stars
fainter than $V\simeq 21$. We ascribe the first problem to the fact that the
synthetic CMDs assume that all the stars of the CMDs of Fig.~\ref{sim} a and b
are actual members of Be29, whilst it may
well be that some objects are not, specially on the brighter parts where the
field star sequence intersects the cluster sequences. 
Indeed, the spread around the subgiant and
red giant branches does suggest the presence of some residual contamination. 
This explanation makes the second problem more significant, since 
contaminating objects may  be present also in the fainter regions of the
empirical CMD, thus making the intrinsic LF of Be 29 lower than the shown one.
We suggest that the discrepancy between synthetic and empirical LF at faint
magnitudes be due to evaporation of low mass stars from the cluster.

\section{Summary and discussion}

From comparison of the observed CMD's to synthetic ones  we obtain the
following results:

i) of the three different sets of evolutionary tracks used, the best
reproductions come from the BBC and FST ones, i.e. the ones taking overshooting
from convective regions into account;

ii) the best solutions are obtained for abundances lower than solar 
(Z=0.02 is excluded), with formal metallicity [Fe/H] $\simeq$ --0.5 or --0.7.
This is higher than estimated by K94, and in very good agreement with the
results of Bragaglia et al. (2004: [Fe/H] = $-0.74 \pm 0.18$) obtained in a 
completely independent way;

iii) the age of Be 29 is found to be around 3.5 Gyr (best synthetic CMDs
provide values from 3.4 to 3.8 Gyr), and this is in reasonable agreement with
the value given by K94;

iv) the best reddening value is around 0.1 mag (\ebv = 0.10 to 0.13), that
well compares to the Schlegel et al. (1998) value of 0.093. The strong
disagreement with the 0.21 value adopted by K94 explains the different 
distance moduli;

v) the best solutions for the absolute distance modulus of Be 29 give  \mmm =
15.6 or 15.8. This puts the cluster at least as far as, or even farther than,
Saurer A from the Galactic centre, with R$_{\rm GC}$ = 21.4 or 22.6 kpc,
thus reinforcing its place of most distant in the still scarce family of very
far and  old open clusters known in the Milky Way.

Be 29 is then an old, relatively metal-poor, distant open cluster. This makes 
it nicely consistent with what is expected from simple chemical evolution 
arguments, predicting that old stars are more metal poor than younger stars
formed in the same place and that the metallicity of coeval objects decreases
with increasing distance from the Galactic centre.  
The precise and detailed metal abundance of Be 29, derived from
high resolution spectroscopy and fine abundance analysis, would be very
valuable in defining the metallicity gradient along the entire disc.

The conclusion that Be 29 appears as a well behaved object in the framework of
Galactic chemical evolution is challenged by a recent suggestion that this
cluster actually is not of truly Galactic origin, but is somehow connected to
the interaction between the Galaxy and its merging satellite Canis Major.  
Frinchaboy et al. (2004) suggest  that the old and far  open clusters, 
specially those located in the two Galactic quadrants towards
the anticentre, lie along a string-like configuration, and follow quite
accurately the distribution of M stars defining the Galactic anticentre stellar
structure (GASS). The coincidence exists both in position (see their fig. 1) 
and in the
l-v$_{GSR}$ plane (their fig. 2), where v$_{GSR}$ is the radial velocity
corrected to the Galactic centre. They lack velocity information on Be 29, but
we have it (Bragaglia et al. 2004), and it puts the cluster right in the GASS
alignment. 
If Be 29 was really born in the dwarf galaxy Canis Major, or formed by the
interaction of the satellite gas  with the  Milky Way gaseous disc component,
its study may be useful to characterize the star formation history of
disturbed regions of our Galaxy. 

Our current data don't allow us to discriminate between the Galactic or external
origin of Be 29. The circumstance that its CMD doesn't show evidence of 
contaminants other than the {\it normal} Galactic disc stars predicted by the 
Besan\c con models seems to suggest that no striking components are 
present in that region. However, our field of view is likely to be too small to
detect this kind of features. We are thus left with the intriguing question of
whether or not the consistency of the properties of Be 29 with the Galactic
age-metallicity relation and metallicity gradient should be taken as a
confirmation of standard Galactic evolution theories.

\bigskip\bigskip\noindent
ACKNOWLEDGEMENTS

This work is based on observations collected at the European Southern 
Observatory, Chile.
We warmly thank P. Montegriffo for his expert assistance with the  data
reduction and the use of his software, and M. Bellazzini for useful
discussions.
We thank Franca D'Antona and Paolo Ventura for providing their stellar
evolution models prior to publication. The bulk of the numerical code for 
CMD simulations was originally provided by Laura Greggio. L.D.F. has been 
funded by a contract at the Osservatorio Astronomico di Bologna.
This work was partly supported by MURST-Cofin98, under the project ``Stellar
Evolution'' and MIUR-Cofin2000 under the project ''Stellar observables of
cosmological relevance''.
This research has made use of the Simbad database, operated at CDS,
Strasbourg, France and of the BDA database, maintained by J.C. Mermilliod.

\end{document}